\documentclass[11pt,draftcls,onecolumn]{IEEEtran}
\usepackage{graphicx}
\usepackage{caption}
\usepackage{subfigure}
\usepackage{amsmath}
\usepackage{amssymb}
\usepackage{bm}
\usepackage{color}
\usepackage{url}
\usepackage{epstopdf}
\usepackage{subfloat}

\graphicspath{{./fig/}}

\newcommand{\qg}{{\bf g}}

\newcommand{\BS}{{BS }}
\newcommand{\User}{{LU }}
\newcommand{\ED}{{ED }}

\newcommand{\EDs}{{ED's }}

\begin{document}
\title{Physical Layer Security for Massive MIMO: An Overview on Passive Eavesdropping and Active Attacks}
 %\markboth{\textit{A Revision Submitted to  The IEEE  Communications Magazine}}
 {}

\author{D\v{z}evdan Kapetanovi\'{c}, Gan Zheng and Fredrik Rusek
\thanks{D\v{z}evdan Kapetanovi\'{c} did this work while at the Interdisciplinary Center for Security, Reliability and Trust (SnT),
University of Luxembourg, Luxembourg, E-mail: {\sf dzevdan.kapetanovic@gmail.com.}}
 \thanks{Gan Zheng is with School of Computer Science and Electronic Engineering, University of Essex, UK, E-mail: {\sf ganzheng@essex.ac.uk.}}
 \thanks{Fredrik Rusek is with Dept. of Electrical and Information Technology, Lund University, Lund, Sweden,  E-mail: {\sf fredrik.rusek@eit.lth.se.}}
 }

\date{\today}
 \maketitle

\begin{abstract}
 This article discusses opportunities and challenges of physical layer security integration in massive multiple-input
 multiple-output (MaMIMO) systems. Specifically, we first show that MaMIMO itself is
 robust against passive eavesdropping attacks. We then review a pilot contamination
scheme which actively attacks the channel estimation process. This pilot contamination attack is not only
 dramatically reducing the achievable secrecy capacity but is also difficult to
 detect. We proceed by reviewing some methods from literature that detect active attacks on MaMIMO.
 The last part of the paper surveys the open research problems that we believe are the most important to address in the future and give a few promising directions of research to solve them.
\end{abstract}

\newpage
\section{Introduction}
During the recent past the field of massive multiple-input
multiple-output (MaMIMO) systems has quickly emerged as one of the most
promising techniques to boost the system throughput of emerging and
future communication systems. In MaMIMO, the vision is to equip the
base station (BS) with an antenna array comprising 100+ antenna
elements  and a large number of independent transceiver chains. An
impressive amount of research on the topic has been conducted, and
MaMIMO will be integrated in the upcoming 5G standard \cite{MaMIMO_Larsson,Rusek}.
The advantages of MaMIMO are manifold, and to give a few we mention
(i) An array gain corresponding to the number of BS antenna
elements, (ii) A channel hardening effect, rendering stable and
predictable channel conditions to users, (iii) Nearly orthogonal
channels from the BS to the users, (iv) Simple signal processing at
both the BS and at the users.

Another advantage of MaMIMO, yet not widely recognized, is that the
potential of physical layer security (PLS) against passive
eavesdropping attacks is increased dramatically. PLS is based on the
following result: Consider a Gaussian wiretap channel \cite{MISOME}
where a BS communicates with a legitime user (LU) in the presence of
a passive eavesdropper (ED). Assume that the Shannon capacities from
the BS to the LU and the ED are $\mathcal{C}_{\mathrm{LU}}$ and
$\mathcal{C}_{\mathrm{ED}}$, respectively. Then, the \emph{secrecy
capacity} between the BS and the LU is
$\mathcal{C}_{\mathrm{SC}}=\max\{\mathcal{C}_{\mathrm{LU}}
-\mathcal{C}_{\mathrm{ED}},0\}$. This rate can be transmitted
reliably and securely without any use of a formal crypto system.
In conventional MIMO systems, the two capacities
$\mathcal{C}_{\mathrm{LU}}$ and $\mathcal{C}_{\mathrm{ED}}$ are of
similar order of magnitude, rendering a fairly small secrecy
capacity. With MaMIMO and passive eavesdropping, the situation changes dramatically. With
standard time-division duplex (TDD) mode MaMIMO operations, and due
to (i) -- (iii) above, the received signal power at the LU is
several orders of magnitude larger than the received signal power at
the ED. This generates a situation where the secrecy capacity is
nearly the full capacity to the LU, i.e.,
$\mathcal{C}_{\mathrm{SC}}\approx\mathcal{C}_{\mathrm{LU}}$
\cite{MaMIMO_passive}. Altogether, MaMIMO enables excellent PLS,
without any extra effort.

There are of course countermeasures that can be taken by the ED.
First of all, it could place itself physically close to the   LU so
that the channels to the LU and the ED are highly correlated. In
this case, (iii) does not longer hold true, and the secrecy capacity
may be compromised. Another effective strategy that the ED can adopt
is to exploit the weakness of the channel estimation phase in
MaMIMO. By switching from a passive to an active mode, the ED can
pretend to be the LU and send a pilot sequence of his own -- this is
the so called pilot contamination attack. The BS will then beamform
signal power to the ED instead of the LU. To detect such a stealthy
attack is challenging since there is a normal amount of pilot
contamination already present in MaMIMO systems.

Active ED attacks are by no means unique to MaMIMO. On the contrary, jamming the BS is a well researched attack in conventional MIMO. See for example \cite{Vulnerabilities,Clancy-ETT13} for two recent papers where the ED attacks the channel estimation phase. The ED can also combine passive eavesdropping and
active jamming attacks. The strategies for countering such an attack in
conventional MIMO channels are discussed in \cite{MS10}. A
 game-theoretic approach is taken in \cite{ZSHPB11} to deal with
combination of passive and active attacks. However, none of the
aforementioned papers explicitly dealt  with MaMIMO systems and the inefficiency of passive eavesdropping in MaMIMO was not observed.

Although MaMIMO has received huge attention, the existing literature
on the combination of PLS and MaMIMO is scarce. This paper will
survey the opportunities that MaMIMO may bring for making PLS a
reality, as well as discussing problems that must be tackled in the
future. The paper sets off by discussing the benefits that MaMIMO
brings to PLS in the presence of a passive ED. We then highlight that
active attacks are more likely to occur for MaMIMO. Three detection methods to deal with active attacks are then
briefly reviewed. We then outline important open problems to
address. The paper is concluded by a discussion of a few promising
directions of future research. We believe that there are many
interesting aspects of PLS integration in MaMIMO to be researched
and hope that this survey will attract the attention of the research
community to this exciting and open field.

\section{Passive and Active Eavesdropping Attacks}
For conceptual simplicity, we consider a single cell with an
$M$-antenna BS, one single-antenna LU, and one single-antenna ED.
The uplink channels from the \User and the \ED to the \BS are
denoted as $\qg_{\mathrm{LU}}$ and $\qg_{\mathrm{ED}}$,
respectively. We assume a TDD system where channel reciprocity
holds, and the corresponding downlink channels are
$\qg_{\mathrm{LU}}^{\mathrm{T}}$ and
$\qg_{\mathrm{ED}}^{\mathrm{T}}$ where $(\cdot)^{\mathrm{T}}$
denotes transpose operation. Standard TDD MaMIMO involves two
phases: (i) The LU transmits a training symbol to the BS in the
uplink, and (ii) Relying on channel reciprocity, the BS performs
channel estimation and beamforms the signal to the LU in the
downlink using the uplink channel estimation $\hat\qg_{\mathrm{U}}$
with proper scaling.

The ED aims to overhear the communication from the BS to the LU  while at the same time being undetected. To this
end, the ED can launch either a passive attack or an active attack, which
will be reviewed below.

\subsection{Passive Eavesdropping Attack}
\begin{figure}[ht]
\centering %\vspace{-10mm}
\subfigure{%
\includegraphics[width=0.48\textwidth]{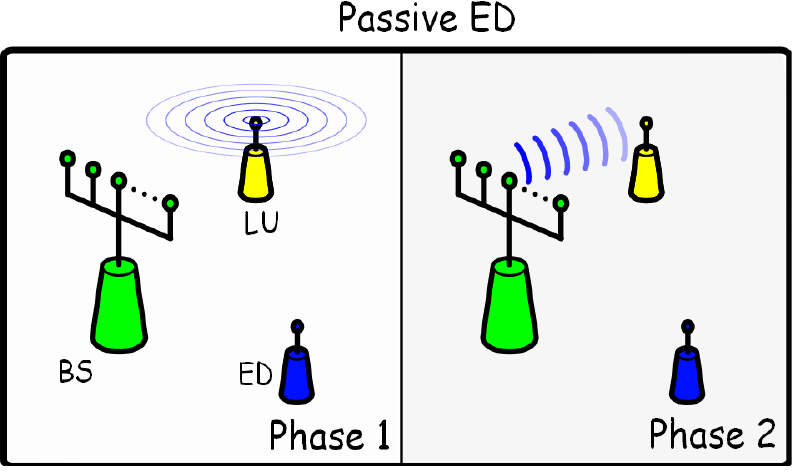}}
\subfigure{%
%\centering \vspace{0mm}
\raisebox{-5mm}{\includegraphics[width=0.49\textwidth]{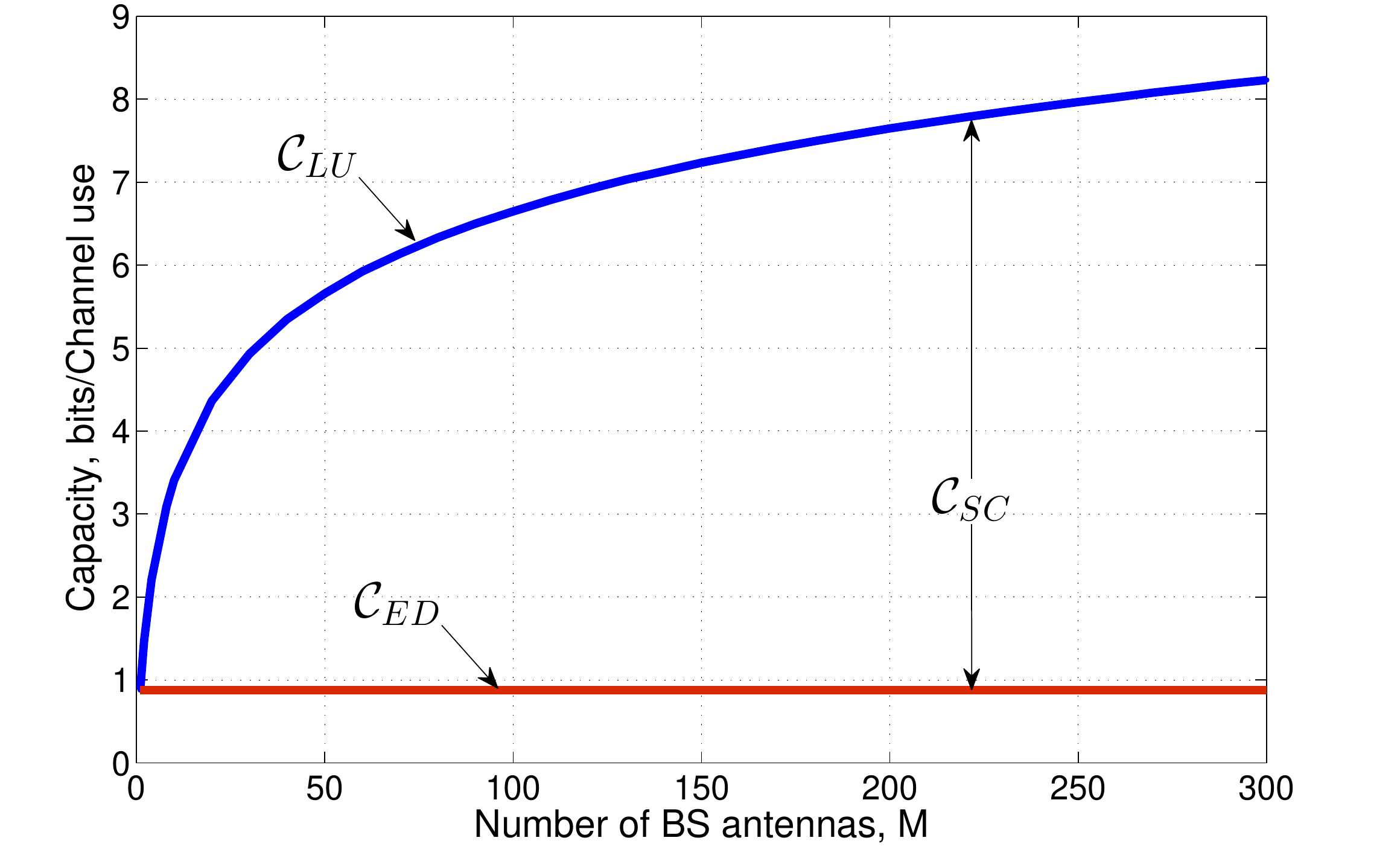}}}
\caption{Left:  A single-antenna passive ED in a single cell containing a single-antenna LU. Right: Example of secrecy
capacity (length of the vertical line).  The ED's capacity
 $\mathcal{C}_{\mathrm{ED}}$ becomes independent of $M$. The secrecy
 capacity increases with $M$.}\label{fig:caps}
\end{figure}

Let us now discuss the passive attack within a MaMIMO context as
shown in Fig. \ref{fig:caps}. The key observation here is that the
presence of a  passive ED is not at all affecting the beamforming at
the BS and has an negligible effect on the secrecy capacity.
Intuitively, this is because MaMIMO has the capability to focus the
transmission energy in the direction of the LU. This implies that
the received signal strength at the ED is much less than that at the
LU. In Fig. \ref{fig:caps}, the resulting ergodic capacities
$\mathcal{C}_{\mathrm{LU}}$ and $\mathcal{C}_{\mathrm{ED}}$ are
shown as functions of the number of BS antennas $M$ with perfect
channel estimation.
%The ergodic capacities are computed for perfect channel estimation at both the LU and the ED.
It is assumed that $\qg_{\mathrm{LU}}$
and $\qg_{\mathrm{ED}}$ are independent and identically
distributed (i.i.d) complex Gaussian vectors with equal mean powers,
and the BS transmit power is normalized to 0 dB.

As can be seen from Fig. \ref{fig:caps}, the ED's capacity remains
the same as $M$ increases; this is so since the BS does not beamform
in the ED's direction. The capacity to the LU, however, is greatly
increased for large values of $M$. In other words, the secrecy
capacity is about half of the LU
capacity with conventional MIMO ($M\approx 2-8$), while it
constitutes more than 85\% of $\mathcal{C}_{\mathrm{LU}}$ already at
$M=100$. From this simple example, we can see that MaMIMO has
excellent potential for integration of PLS.

\subsection{An active attack on the channel estimation}
\label{sec:activeattack}
 The resilience of MaMIMO against the passive attack is based on the
 assumption that the uplink channel estimation $\hat\qg_{\mathrm{LU}}$ is independent
 of the ED's channel $\qg_{\mathrm{ED}}$. This motivates the \ED
 to design active attacks on the channel estimation process to
 influence  the BS's beamforming design. Next we describe
 such an attack based on the pilot contamination scheme in \cite{Attack_Zhou}.

\begin{figure}[ht]
\centering %\vspace{-10mm}
\subfigure{%
\raisebox{-15mm}{\includegraphics[width=0.49\textwidth]{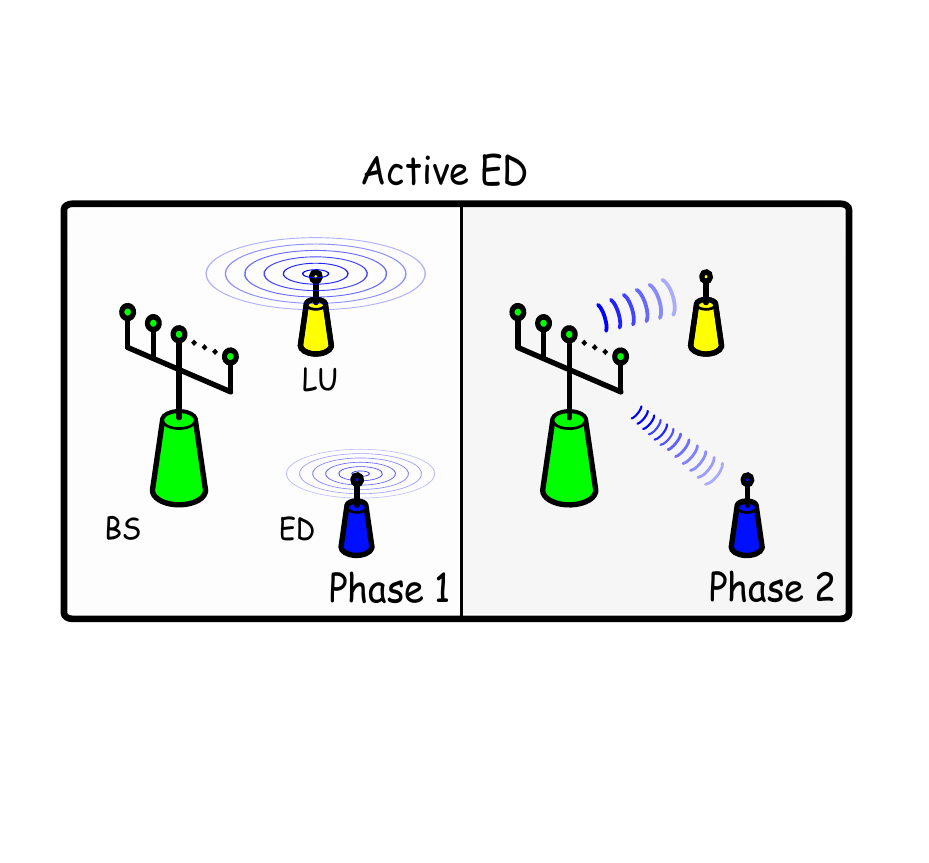}}}
\subfigure{%
\includegraphics[width=0.49\textwidth]{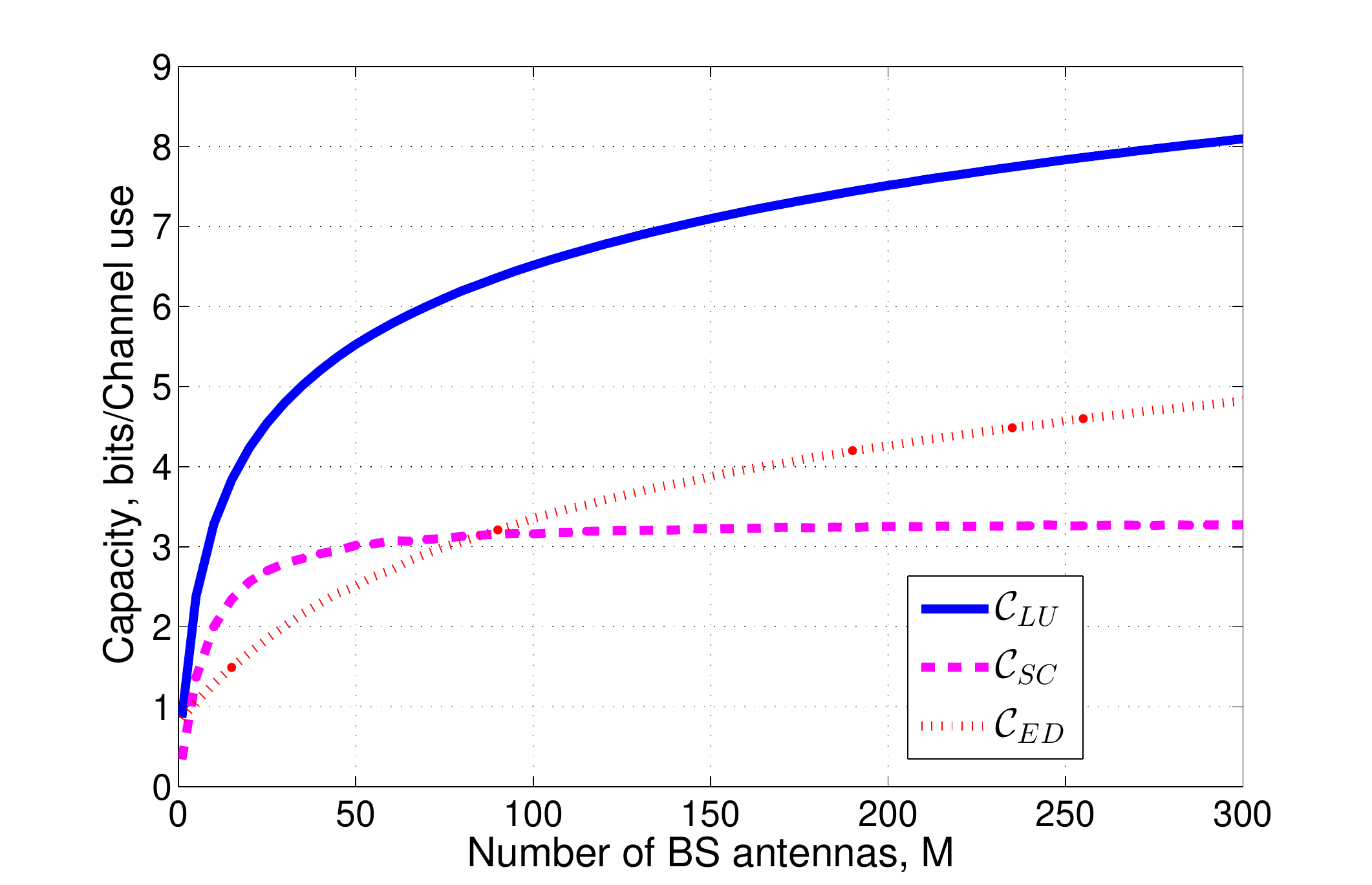}}
\vspace{-20mm}\caption{Left:  Active attack on the channel estimation. Right: The
resulting channel capacities and the secrecy capacity. The ED's
training power is 10 dB weaker than the LU's.}\label{fig:attack:CSI}
\end{figure}

 As illustrated in Fig. \ref{fig:attack:CSI}, during the
 uplink channel estimation, the \User transmits a pilot symbol to the BS. At the same time, the \ED launches the attack  by sending
 another pilot symbol. In the worst case the \ED is synchronized with the legitimate transmission
 and this is possible by  overhearing the signaling exchange  between the \BS and
 the LU.

 The consequence of this attack, if left undetected, is that the promising PLS benefits of MaMIMO are lost.
 The difference from the passive attack is that the channel estimate $\hat \qg_{\mathrm{LU}}$ becomes correlated with the \EDs channel $\qg_{\mathrm{ED}}$, and consequently the
 equivalent channel for the \ED also improves as $M$
 increases.  Even worse, if the ED uses higher training power, it dominates the training phase and
the secrecy capacity may become zero.

 For the same settings as in Fig. 1, we plot the ergodic channel capacities and the ergodic secrecy capacity   in Fig. \ref{fig:attack:CSI}.
 The signal sent by the \ED is 10dB weaker than the training signal sent by the LU.
 %(When the ED and the LU use the same power, the ergodic secrecy capacity is zero regardless of $M$.)
 As can be seen, both the ED's  and the LU's channel capacities increase with M,
 however, the secrecy capacity remains constant for
 $M>50$.

 Although the  situation is similar to the well-known pilot
 contamination  problem  in multi-cell systems \cite{contamination}, a notable difference is that
the \ED is out of control and   therefore existing schemes for
reducing pilot contamination in MaMIMO cannot be applied. In the
remainder of this paper we survey recent progress to detect the
active attack and discuss possible future directions of research.

\section{Detection Schemes} \label{sec:detschemes}
As described in Section \ref{sec:activeattack}, detection of an active ED
is crucial for secure MaMIMO communication. Now we will present arguments that show how peculiar and different ED detection is in MaMIMO systems.

Consider a detection scheme applied by the BS during uplink packet transmission, that is based on successful packet reception. In systems such as LTE and WLAN, the uplink packet contains channel estimation pilots. If the ED attacks these pilots, one could argue that this would result in decoding errors and thus packet loss (due to a bad channel estimate). Indeed, this is what happens in conventional MIMO systems with few antennas (depending on the robustness of the used modulation and coding scheme, of course). Therefore, the BS would suspect presence of strong interference, either coming from an ED or another user, and could take actions based on this.  However, in MaMIMO, the erroneous channel estimate does not typically result in a decoding error, assuming that ED's channel is uncorrelated with LU's channel. Hence, in contrast to a conventional MIMO system, a successful packet reception does not imply absence of an ED in a MaMIMO system. As we have argued, if this erroneous channel estimate is left undetected and arises from an active ED, it can have a detrimental effect on the secrecy capacity if used in the subsequent downlink phase. Hence, the importance of effective detection schemes during channel estimation.

An appealing and
conceptually simple detection strategy that can be used during channel estimation, which we will also argue against,
is to let the BS first estimate the mean power or large scale fading
$\beta$ of the estimated uplink channel vector
$\hat{\mathbf{g}}_{\mathrm{LU}}$, i.e.,
$\beta=\mathbb{E}{[\|\hat{\mathbf{g}}_{\mathrm{LU}}\|^2]}$. The
value $\beta$ changes slowly over time/frequency so that a good
estimation of it is feasible. The instantaneous received energy of a
pilot observation at the BS converges to $\beta P + N_0$ as $M$
grows large, where $P$ is the power of the pilot symbol and $N_0$ is
the noise density at the BS. Thus, if $\beta$ and $N_0$ are known,
the BS can compare the instantaneous received energy with $\beta
P + N_0$. In the absence of an active \ED the two quantities are
close to each other, while an active ED is detected  if the
instantaneous received energy is much larger than $\beta P + N_0$.

Unfortunately, there is a simple countermeasure that the ED can
take. Since the value of $\beta$ changes slowly, the ED
can adapt its transmit power  in order to sabotage the
estimation of $\beta$ at the BS. The ED starts transmitting at low
power, and increases the power over several coherence intervals of
$\beta$. The ED is thereby emulating the natural change of the
channel propagation environment, and the BS can not distinguish the
increased received power from a natural channel quality improvement.
The lesson learned from this example is that detection methods
should not solely depend on the large scale fading parameter $\beta$
and should preferably work without the knowledge of $\beta$. This
section will describe two different schemes (one with two flavours) that can  effectively detect such
an attack in MaMIMO systems without the knowledge of $\beta$. Both schemes share two important features: simplicity and effectiveness, which are due to MaMIMO. Although ED's presence during the channel estimation phase removes the expected MaMIMO gain in secrecy capacity, as demonstrated in section \ref{sec:activeattack}, MaMIMO instead enables simple and effective schemes for detection of the ED.

\subsection{Detection Scheme 1: LU transmits random pilot symbols} \label{sec:det:1}
 This scheme exploits controlled randomness in transmitting ``random'' pilots  to detect an active ED.
 As first proposed in \cite{PIMRC_13}, the LU transmits a sequence of random
  phase-shift keying  (PSK) symbols, which forms the key to detecting
 the ED at the BS.  Below we discuss two variations of this scheme.

\subsubsection{Scheme 1a}
Two pilot symbols $p_1$ and $p_2$ transmitted by the LU are
chosen independently from an $N$-PSK constellation. The BS receives the two pilot
signals $\mathbf{y}_1$ and $\mathbf{y}_2$. The
BS now forms the detection statistic $z$ as the phase of
$\mathbf{y}_1^{\mathrm{H}}\mathbf{y}_2$, where
$(.)^H$ is conjugate transpose. Three scenarios are possible:
\begin{enumerate}
\item The ED is absent during both transmissions. In this case, the phase of $z$ can be shown to converge to a phase of a valid PSK symbol as the number of antennas grows large.
\item The ED is present in both slots. As the number of antennas grow large, the phase of $z$ converges with probability $1 - 1/N$ to a number that is not a valid phase of an $N$-PSK symbol.
\item The ED is present in only one of the slots. In this case,  $z$ still converges to a valid PSK phase. However,
the received power at the BS will be biased towards a larger value during the slot where the ED is present. Hence,
 one can form the ratio $q = \|\mathbf{y}_1\|^2/\|\mathbf{y}_2\|^2$. If $q < \gamma_1$ or $q > \gamma_2$, for some thresholds
 $\gamma_1,\gamma_2$, then it is decided that the ED is present.
\end{enumerate}
From 2), we know that as the antenna number $M$ grows large, the probability of detection converges to $1-1/N$. Thus, it can be made arbitrarily close to 1 by increasing the alphabet size $N$ while the false alarm probability converges to 0 due to 1). In order to use a large value of $N$, $M$ must be quite large ($M > 200$).  Note that this simple scheme is effective due to MaMIMO, which provides convergence of the scalar product to different values depending on ED's presence. Moreover, for large $M$, this scheme is robust to knowledge of the noise power $N_0$. Namely, the noise is averaged out in the scalar product between the received signals when $M$ is large.  %as the energy test 3) can be removed [{\rl under which condition, 3) is unnecessary?}].%Figure \ref{fig:sch1a} shows this behaviour for cases 1 and 2. Notice that the performance is grossly improved if more than 2 symbols are used.

\subsubsection{Scheme 1b}
The second random pilot scheme provides improvements to the above scheme in case of three or more observations. Given $L$ observations $\mathbf{y}_1,\ldots,\mathbf{y}_L$, form the matrices
\begin{eqnarray}
\label{eq:Yeq}
\mathbf{Y} & = & [\mathbf{y}_1,\ldots,\mathbf{y}_L], \nonumber \\
\mathbf{R} & = & \frac{\mathbf{Y}^H\mathbf{Y}}{M} - N_0\mathbf{I},
\end{eqnarray}
where $\mathbf{I}$ is an $L\times L$ identity matrix. When $M$
becomes large, $\mathbf{R}$ converges to a rank-one matrix if ED is
absent, otherwise it converges to a full rank matrix with
probability $1 - 1/N$. Based on this observation, the following
detection rule is applied: if
$\lambda_1\{\mathbf{R}\}/\lambda_2\{\mathbf{R}\}$ is greater than
some threshold, ED is absent; otherwise it is present, where
$\lambda_1\{\mathbf{R}\}$ and $\lambda_2\{\mathbf{R}\}$ are the
largest and the second largest eigenvalues of $\mathbf{R}$.

We will later see in section \ref{sec:comp} that this scheme provides significant performance enhancement with only four observations. The scheme also takes care of the case when the ED is not present in all $L$ slots. Again, noteworthy is the simplicity of this scheme, which is due to MaMIMO since it enables convergence of $\mathbf{R}$ to matrices of different rank depending on ED's presence. Although performing significantly better than scheme 1a, scheme 1b requires a good estimate of $N_0$, since it is used to construct $\mathbf{R}$.

\subsection{Detection Scheme 2: Cooperative Detection Scheme}
\label{seq:copsch} The above random detection incurs the overhead of
transmitting additional random sequences. We next briefly introduce
a detection method that does not transmit additional pilot symbols
\cite{PIMRC_14}.

As illustrated in Fig. \ref{fig:def4}, upon receiving the training
signal from the LU, the BS can apply a beamformer based
on the received signal, and then transmit a pilot to the LU using
the same beamformer. The beamformer is constructed in such a way
that the received sample at the LU in the absence of an active ED
equals an agreed value (between the BS and the LU) after a scaling with $1/M$.  For simplicity, this agreed value can be taken as 1 in our discussion.
%Notice that since there is noise present in the channel estimate $\hat{\mathbf{g}}_{\mathrm{LU}}$ regardless if there is any active ED or not, the noise density $N_0$ must be known at the BS in order to ensure that the agreed value 1 appears at the LU.
In the case
that there is an active ED, the LU will observe a much smaller
quantity, and this forms our basis to detect the presence of the ED.

As before, simple beamforming is effective due to MaMIMO. Similar to the random pilot scheme 1b, the cooperative scheme also requires a decent estimate of $N_0$ for proper performance.

\subsection{Comparison of Detection Schemes}
\label{sec:comp} We discuss  the pros and cons of the introduced
detection schemes and show a comparison of detection performance in
Fig. \ref{fig:comp}  with 200 antennas at the BS. The channel
vectors $\qg_{\mathrm{LU}}$ and $\qg_{\mathrm{ED}}$ are independent
with each element drawn from a complex Gaussian distribution with
mean 0 and variance 1. The pilot powers from the LU and the ED are also equal. For the random pilot
schemes, we use a QPSK alphabet. For the cooperative scheme, it is
assumed that the noise power at the BS equals the noise power at the
LU.

As shown in Fig. \ref{fig:comp}, the cooperative scheme performs
best at moderate to high SNRs. The dotted curve illustrates the
performance of  random pilot scheme 1a. Its performance is inferior
to the cooperative scheme at moderate to high SNRs, but
significantly better at low SNRs. The dashed curve represents
random pilot scheme 1b with four received slots, which exhibits very good
performance compared to the random pilot scheme with two slots.

%Both the random pilot scheme and the cooperative scheme can  be applied to today's wireless systems such as WLAN without large obstacles (as
% will be explained in Section \ref{sec:standards}).
As depicted in Fig. \ref{fig:comp}, the biggest advantage of the
cooperative scheme is its detection performance during the two
exchange intervals. The drawback with the cooperative scheme is that
the ED can cause problems during the whole frame exchange. For
example, the ED might be very close to the LU, and thus contaminate
any packets that the LU receives from the BS. The noise power at the
LU will therefore be significantly higher than the noise power at
the BS, causing a degradation in the detection performance at the
LU.
%In that case, the performance of the cooperative scheme in
%Figure \ref{fig:comp} can easily drop below the random pilot scheme
%1b (and below the random pilot scheme 1a if the power from the ED is
%significantly large).

The advantage of the random pilot schemes are their robustness to
jamming of the LU by the ED, since the detection is performed at the
BS. If the ED would increase its power toward the BS, it would
instead result in yet easier detection of the ED. However, as
illustrated in Fig. \ref{fig:comp}, the random pilot schemes
requires more than two observations of the received signal in order
to have a performance comparable or better than the cooperative
scheme.

\section{Further Discussions and Future Directions} \label{sec:furtherdisc}
 In this section, we discuss
 limitations of the solutions introduced above and give promising future directions.

\subsection{Limitations}
The reason why only the active attack is effective in MaMIMO  is
that the channels to the LU and the ED are assumed to be largely
uncorrelated. If the channels were correlated, or in the worst case
the same, then there is no need for the ED to be active. This means
that if the ED can position itself   such   that its channel to the
BS is highly correlated with the channel from the LU to the BS, then
the ED can be passive. Furthermore, in line-of-sight scenarios the
beamforming at the BS is directional. If the ED is at the very same
angle-of-departure from the BS, the ED will receive a highly
correlated signal although it may not be very close. We point out
that this problem is alleviated in 3D beamforming scenarios. These
considerations are important  to take into account in further work
to delineate suitable scenarios for combining PLS with MaMIMO.

\subsection{Detection of Active Attacks in Multi-user and Multi-cell Systems}
We have so far been dealing solely with a single cell and a single
user. The case of a multi-user scenario does not alter the situation
much as the users can be allocated orthogonal resources for
transmitting training symbols. Thus, there is virtually no
interaction among users during the training phase. A multi-cell
scenario, on the other hand, brings about radical changes. This is
so since even without any active ED, the received signal at the BS
corresponding to the LU's training signal is being interfered with
by LUs from other cells; this is the so called pilot contamination
problem of massive MIMO \cite{Rusek}. Consequently, in a multi-cell
scenario, the problem of detecting the presence of an active ED
changes into the much more challenging problem of distinguishing an
active ED from LUs in other cells. In a multi-cell scenario, the
detection methods from Section \ref{sec:detschemes} will all fail
since they will detect the presence of the other cells' LUs, but are
not sophisticated enough to distinguish these from an active ED.

The multi-cell active ED detection problem is a wide open research
problem as, to the best of the authors' knowledge, there are no
attempts in the existing literature to deal with it. Although not
much is known, we briefly introduce two potential ways forward that
can be pursued in future research.

\subsubsection{Cooperative BSs}
Cooperative BSs is a technology that has already found its way into
the LTE  standard, both in the form of the Coordinated Multipoint
(CoMP) and as the less complex Network Assisted Interference
Cancellation and Suppression (NAICS). In these architectures, the BSs of different
cells are connected via a backhaul network and exchange information.
Consequently, there is a possibility to let the BSs jointly estimate
the level of LU-induced pilot contamination. Even better, the
cooperative BSs can jointly try to minimize the pilot contamination
in the system. Then the detection methods in Section
\ref{sec:detschemes} can be applied verbatim. If the detection
methods identify a suspiciously high level of pilot contamination,
then the transmission can be terminated. In the case that there was
in fact no active ED, but the LU-induced pilot contamination level
was unusually high, the terminated transmission is not a major
problem as the transmission is not very efficient whenever the pilot
contamination is high.

\subsubsection{Methods based on radio propagation characteristics} \label{radiobased}
In a multi-cell case, pilot contamination is caused by LUs from
other cells.  To be effective, the ED should arguably be located
within the serving cell. The radio propagation characteristics
between users and a MaMIMO BS are today well understood, and a
reasonable working assumption is that the statistical properties of
incoming radio waves to the BS from LUs far away are different from
those coming from a potential ED that is located much closer to the
BS. This is so since signals from users in other cells are typically
being reflected by a few dominant objects in the vicinity of the
users, and these objects may have line-of-sight to the serving cell
BS. Thus, the radio waves' angle-of-arrivals (AoAs) for users in other
cells may be limited to a few possibilities, while the signal from a
close by ED may reach the BS via interacting objects in the vicinity
of the BS. This opens up a possibility to differentiate between
pilot contamination from LUs and pilot contamination from an active
ED.

\subsubsection{The angle-of-arrival database for location-aware users}
A particularly interesting direction of research may be to combine
the outlined approach in  Section \ref{radiobased} with the results
from \cite{Location}. In \cite{Location}, the authors discuss
possibilities of systems where LUs can obtain position estimates of
themselves. If the LUs report their positions and signal-strengths
to a central node, then a database of signal strength as function of
physical position can be constructed. In MaMIMO, this idea can be
extended and utilized to detect active EDs. Rather than building a
database of signal-strengths, a database can contain the
AoAs to the BS from LUs at certain positions. The detection
of an active ED would then comprise the following steps: (i) the
\User reports his position to the \BS, (ii) the \User transmits a
training symbol, (iii) the \BS requests the AoAs for
the particular user position from the database, (iv) if the measured
AoAs for the pilot observation does not match the input
from the database, there is an active ED. So far, the authors are
not aware of any attempt in the literature to build an
AoA database, and future research is needed to
investigate the feasibility of this approach.

Again, if the \ED can get physically close to the \User the measured
AoAs may match those in the database even in the
presence of an active \ED. However, recent measurement campaigns
\cite{Pepe} with a 128 port antenna array and where 8 users are
located within a five-meter diameter circle show that the radio
propagation characteristics are sufficiently different for the 8
users (both in line-of-sight and non-line-of-sight) so that they can
all be spatially separated. This indicates that a MaMIMO system may
very well be capable of distinguishing an \ED from a \User even
when they are physically close.

\subsection{Detection using Learning Mechanism}
Machine learning is a powerful tool that comes natural to mind for detecting active EDs. Machine learning comes in two forms: supervised and unsupervised
learning. In supervised learning one must guarantee that there is no active ED present when training the machine.
When this is not possible to guarantee unsupervised
learning should be used. However, we can foresee a number of problems that arise that must be dealt with.
Firstly, a fairly large amount of training data must be used in machine learning, which may cause an unacceptably high overhead. In MaMIMO this is a particular severe problem as the amount of training may grow with the antenna number.
Secondly, the mobility of users is a problem since the channel
quality and characteristics may vary dramatically over time and space.

Alternatively, device dependent radio-metrics such as frequency and phase shift
differences of radio signals can be used as unique fingerprints to
detect active EDs \cite{Fingerprinting}. This
method is channel-invariant, but extracting the features is a
complicated task. More efforts need to be made to resolve these
problems in order to use machine learning tools for detecting active EDs.

\section{Conclusions}
 MaMIMO has shown great potential to boost system capacity but the combination of PLS and MaMIMO is not well understood yet. In this article, we
 have reviewed both attacks and detection schemes. In particular, we have shown that although passive eavesdropping has little effect on the secrecy
 capacity, an active attack on the channel estimation phase is harmful. We
 have presented three detection schemes that can effectively  identify the active attack.
 The detection of active attacks in  multi-cell and  multi-user systems is especially relevant to the applications of MaMIMO - but is challenging and not much is known today.
 Some promising research directions to this end have been discussed in the paper.

\newpage

\newpage

\newpage

\begin{figure}[ht]
\centering %\vspace{-10mm}
\includegraphics[width=0.86\textwidth]{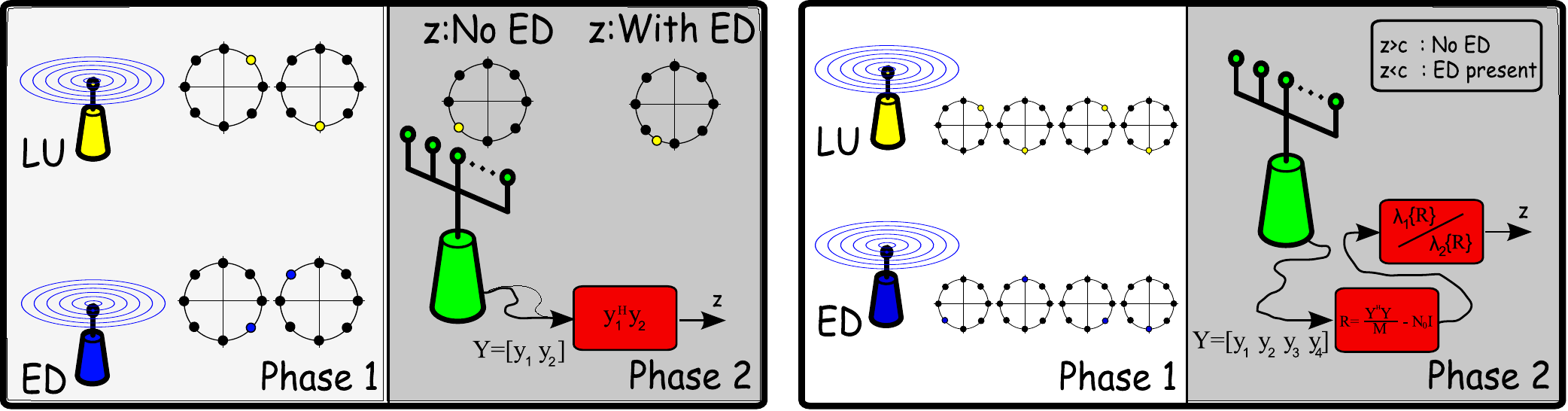}
\caption{Left: Detection scheme 1a.  LU first transmits two
random PSK symbols. After processing at the BS side, the correlation
of the two received signals should (roughly) be a valid PSK symbol
if there is no active ED present. Notice that case 3 is not covered
in the figure. Right: Detection scheme 1b. LU transmits (in this case) four random PSK symbols. The BS constructs a correlation matrix, and performs a test based on the ratio of the two largest eigenvalues of the matrix. \label{fig:def1}}
\end{figure}
\newpage

\begin{figure}[ht]
\centering %\vspace{-10mm}
\includegraphics[width=0.65\textwidth]{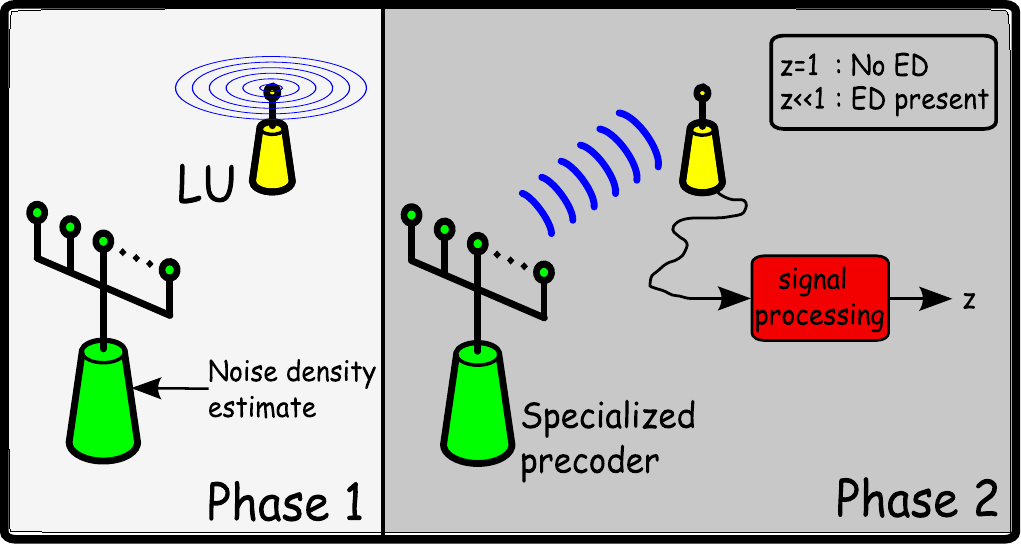}
\caption{Overview of   Detection scheme 2. In phase 1, the BS
performs channel estimation. In the next phase, the BS applies a
specialized beamformer (see \cite{PIMRC_14}) which ensures that the
received signal at the LU after processing becomes 1. If a smaller
quantity is observed, there is an active ED
present.}\label{fig:def4}
\end{figure}
\newpage

\begin{figure}[ht]
\centering %\vspace{-10mm}
\includegraphics[width=0.65\textwidth]{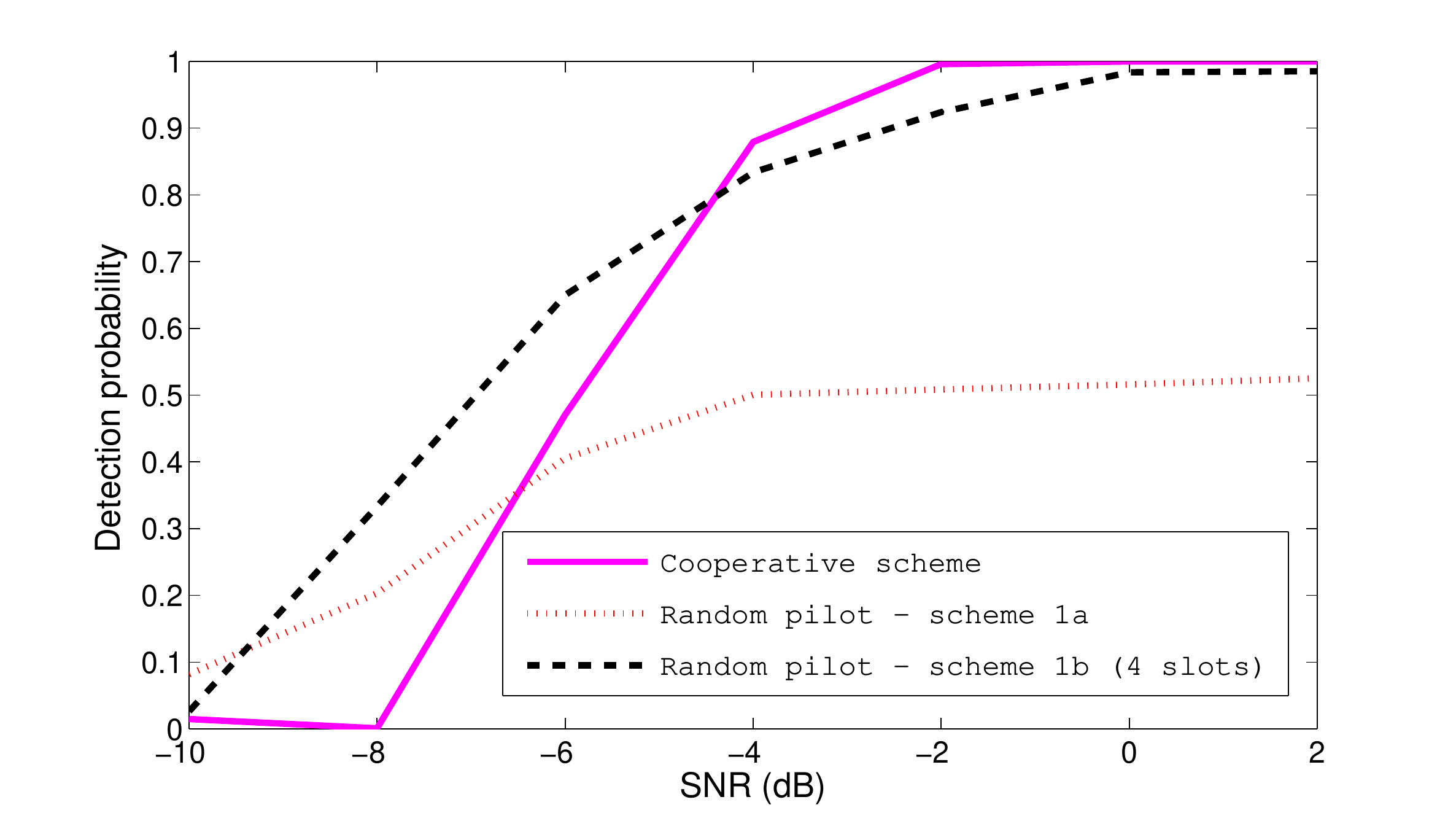}
\caption{Comparison of the discussed detection schemes. The false alarm probability is 1\% in all schemes. The random pilot schemes use random QPSK pilots.}\label{fig:comp}
\end{figure}

 \end{document}